\documentclass[12pt,dvips]{article}
\usepackage{epsfig}
\usepackage{cite}
\usepackage{latexsym}
\usepackage{amstex,amssymb}

\newcommand{\su}{ SU(2)}				
\newcommand{\G}{{\cal G}}			
\newcommand{\fhom}{{\pi}_1}		

\newcommand{\Hil}{{\mathfrak H}}			
\newcommand{\D}{{{\rm Diff}}}
\newcommand{\Df}{{\rm Diff_{F}}}
\newcommand{\Spa}{\Sigma}
\newcommand{\dham}{{\widehat {\mathbb H}}}
\newcommand{\R}{{{\cal  R}}}
\newcommand{\A}{{\cal A}}
\newcommand{\sd}{{\ltimes}}
\newcommand{\re}{{\rm I\!\rm R}}	
\newcommand{\C}{{\mathfrak C}}
\newcommand{\1}{{\bf 1}}

\newcommand{\id}{\equiv}
\newcommand{\Z}{{\mathbb Z}}		
\newcommand{\ho}{{\Hil}^0_q}				
\newcommand{\rd}{{\rm d}}	
\newcommand{\bun}{{\cal H}}
\newcommand{\Ho}{{\bun}^0}
\newcommand{\disk}{{\mathbb D}}
\newcommand{\cAA}{{\mathbb A}}
\newcommand{\cA}{{\mathfrak A}}
\newcommand{\pP}{{\mathbb P}}

\newcommand{\snabla}{\;{/ \! \! \! \!{\nabla}}\;}
\newcommand{\sD}{{/ \! \! \! \!{{\mathfrak D}}}}
\newcommand{\di}{\;{{\sD}}^5(G',\cAA')}
\newcommand{\gdi}{{\snabla}^5}
\newcommand{\dfo}{\;{{\sD}}^4}
\newcommand{\gdfo}{{{\snabla}}^4}
\newcommand{\diD}{\;{\sD}^5(G,\cAA)}
\newcommand{\tsnabla}{{\snabla}^3}
\newcommand{\tnabla}{\nabla^3}
\newcommand{\RP}{{\re}{\pP}^3}

\begin{document}

\begin{flushright}
SU-4240-667 \\
\end{flushright}

\begin{center}
{\large{\bf GLOBAL ANOMALIES IN CANONICAL GRAVITY}}

   Sumati Surya\footnote{ssurya@@iucaa.ernet.in} \\
	{\it Inter-University Centre for Astronomy and Astrophysics,} \\
	{\it Post Bag 4, Ganeshkhind, Pune, India 411007.} \\ 
			{\small and} \\ 
        {\it Department of Physics, Syracuse University,} \\
	{\it Syracuse, N . Y.  13244-1130,  U. S. A.}\\ 

   Sachindeo Vaidya\footnote{sachin@@suhep.syr.edu} \\
	{\it Department of Physics, Syracuse University,} \\
	{\it Syracuse, N . Y.  13244-1130,  U. S. A.} 

\end{center}

\begin{abstract}
	In this note we study the structure of diffeomorphism anomalies in
	$3+1$ canonical gravity coupled to a chiral massless fermion. We
	find that when the spatial manifold $\Spa$ is $S^3$ or a Lens space
	$L(p,q)$ , the first homotopy group of the related diffeomorphism
	group can be nontrivial and hence the question of global anomalies
	becomes relevant.  Here we show that for gravity coupled to $\su$
	chiral fermions, assuming the strong form of the Hatcher
	conjecture, $\su$-induced diffeomorphism anomalies do not occur
	whenever $\Spa$ is $S^3$ or a Lens space.
\end{abstract}

\section{Introduction}
The existence of a global gauge anomaly in a 4-dimensional Euclidean
path-integral picture relies on the non-triviality of the group of large
gauge transformations ${\pi}_0({\G}^4)$.  The $\su$ anomaly can be
attributed, first, to the non-triviality of ${\pi}_0({\G}^4) =
{\pi}_4(\su)=\Z_2$, and then, to the existence of an odd number of zero
modes for an appropriately constructed 5-dimensional Dirac operator
\cite{witten}.  In the Hamiltonian approach studied in \cite{nelalv}, the
$\su$ anomaly comes from a non-trivial Berry-phase picked up by the Dirac
vacuum when it traverses a non-contractible loop, $\gamma \subset
[\gamma]\in \fhom({\G}^3) =\Z_2$. This phase provides a global obstruction
to implementing Gauss' Law, since $\G^3$ no longer has a well-defined action
on the vacuum \cite{segal}.  We note therefore that any gauge theory for
which $\fhom({\G}^3)$ is non-trivial has the potential to be anomalous.

In the case of gravity, the 3-dimensional group of gauge transformations
$\G^3$ is replaced by the diffeomorphism group $\D(\Spa)$ of the spatial
3-manifold $\Sigma$. Unlike gauge transformations, diffeomorphisms are not
fibre-preserving automorphisms of the frame-bundle over $\Sigma$ and the
relation ${\pi}_i({\G}^n)={\pi}_{n+i}(G)$ between the homotopy groups of
the gauge group $G$ and the group of gauge transformations in $n$
dimensions ${\G}^n$, is not valid.  Since $\D(\Spa)$ is not known, even
up-to homotopy for a generic $\Sigma$, it is not even possible to ask
whether such a theory has the {\it potential} to be anomalous, let alone
establishing one way or the other the existence of such an anomaly. Our
attention to this problem was therefore first excited by noticing that work
had been done in precisely this direction in \cite{witt} for $\Sigma$, a
spherical space.

Using Hatcher's conjecture for spherical spaces \footnote{This
conjecture states that $\D(\Spa)$ is of the same homotopy type as the
isometry group of $\Spa$ \cite{hatcher1}.} and the results of \cite{witt},
it can easily be see that $\fhom(\D(\Spa))$ is non-trivial for Lens spaces.
Hence we see that the {\it possibility} of a global anomaly does exist in
$3+1$ quantum gravity coupled to a massless chiral fermion, whenever the
spatial manifold is a Lens space.  We combine this result with the methods
of \cite{witten,nelalv} to examine whether such anomalies exist or not.  We
are constrained for technical reasons to a theory coupled to an $\su$ gauge
field. Our work shows that such theories in fact do not possess $\su$ gauge
induced diffeomorphism anomalies.

The importance of chiral fermions in a fundamental theory is obvious and
the existence of an anomaly is a serious indication that the theory is
inconsistent with the known physical world. If indeed a spatial topology
$\Spa$ exists for which the associated theory is anomalous, one can
conclude that this topology must be ``forbidden''. In other words it would
provide a ``selection rule'' for spatial topology at such energies.  It is
in this spirit that we wish to pose our question.  Of course these
considerations would become inconsequential at energies at which the
``frozen'' topology picture breaks down, but presumably it is still valid
in an effective sense. The presence of the ``external'' $\su$ field as we
mentioned earlier, is only a technical crutch at this point and its
physical importance is not clear. We hope to extend our analysis to the
pure gravity case in the near future.

\section{Global Anomalies and $3+1$ Gravity}
In the standard canonical approach to quantum gravity, spacetime has the
topology $\Spa \times \re$ where the topology of space is ``frozen''.  If
$\R$ denotes the space of all Riemannian metrics on $\Spa$, and $\D$ the
group of all diffeomorphisms of $\Spa$, then the true configuration space
for pure gravity is the space of 3-geometries, $\R/ \D$. Thus $\D$ acts
as the group of gauge transformations for such a theory. If $\Spa$ is
closed, $\D$ contains non-pointed diffeomorphisms that do not act freely on
$\R$. Thus $\R/ \D$ does not have a manifold structure, but is an orbifold.
Orbifold-quantization may be dealt with by removing the isolated singular
points of $\R/ \D$and using appropriate self-adjoint extensions of the
quantum operators (see for example \cite{chaerc} and references
therein). On the other hand, if $\Spa$ is thought of as a one-point
compactification of a manifold that is asymptotically like ${\re}^3$ then
the relevant group is the group of ``frame'' fixing diffeomorphisms, $\Df
\subset \D$ which does act freely on $\R$ so that $\R/\D$ is a manifold. In
what follows we only consider the former case with $\Spa$ closed. Problems
associated with orbifold quantisation do not affect our present analysis,
however, since we are concerned with a more basic question: does
there exist a  well defined action of $\D$ on $\R$ at all, free or otherwise.

The importance of $\D$ having a well defined action on $\R$ becomes clear
when we consider the momentum constraint (Gauss law
constraint). Implementing this constraint on the physical states by the
Dirac procedure makes them invariant under ${\D}_0$, the identity connected
subgroup of $\D$. However, if ${\D}_0$ does not have a well defined action
on $\R$ then it cannot have a well-defined induced action on the states
$\psi: \R \rightarrow {\C}$ and this provides an obstruction to
implementing the constraint \cite{segal}.  In the case of pure gravity,
such a problem does not exist since there is a well defined action of
${\D}_0$ on $\R$ \cite{bamalfi}.  However, when gravity is coupled to an
odd number of chiral fermion fields, one needs to re-examine this question.

Using the ideas presented in\cite{nelalv}, one can carry out the
quantization of gravity coupled to chiral fermions in two stages.  First,
one quantizes the fermion field in a background metric, and second, the
gravitational field. At the end of the first step, for every generic $q \in
\R$, there exists a well-defined Dirac Fock space, ${\Hil}_q$. The Fock
bundle $\bun$ over $\R$ can then be constructed from these 
${\Hil}_q$. The Fock vacuum $\ho$ (defined as the state in Fock space for
which all the negative energy states are filled) will not be well defined
at all points of $\R$, because of the existence of zero energy
eigenvalues. It is the existence of such (non-generic) points that allows
one to check for anomalies.

To complete the quantization, the momentum constraint needs to be
implemented and ${\D}_0$ must have a well defined action on $\Ho$, the
vacuum bundle over $\R$.  Now, the spectrum of the Dirac Hamiltonian
${\dham}_q$ at any generic point $q$ in $\R$ is the same at all $q'$
related to $q$ by a diffeomorphism. However, ${\Hil}^0_{q'}$ need not be
the same as ${\ho}$ for all such $q'$. In particular, if $\D$ is not simply
connected, the Fock bundle may ``twist'' along a non-trivial element of
$\fhom(\D)$. Since $\dham$ is a real operator, one can always choose a real
Hilbert space on $\R$ (c.f. \cite{atisin} for meaning of real operator) and
hence the phase reduces to $\pm 1$. The existence of a twist implies that
${\D}_0$ does not have a well defined action on $\Ho$ which in turn points
to a global obstruction in implementing the momentum constraint.

Clearly, if $\D$ {\it is} simply connected, by continuity arguments, such a
twisting is not in principle possible, and one need not worry about the
possibility of an anomaly. Thus, the first question we need to ask is
whether $\D$ is simply connected or not. Below we see that the first
homotopy group for both the full diffeomorphism group and the frame fixing
subgroup is nontrivial for several Lens spaces.

\subsection{$\fhom(\D)$ for Lens Spaces}
The Hatcher conjecture,  ${\pi}_n(Isom)\simeq{\pi}_n(\D)$, holds for
both $S^3$ and $\RP$ \cite{hatcher1,hatcher2,ivanov}. For all other Lens
spaces, a weaker form of the conjecture holds, i.e,
${\pi}_0(Isom)\simeq{\pi}_0(\D)$ and the number of generators of
${\pi}_1(Isom)$ is equal to the number of generators of ${\pi}_1(\D)$
\cite{rubinstein,rubbir,hodrub}. {\it Assuming} the strong form of the
conjecture  for all Lens spaces $L(p,q)$, we simply read off $\fhom(\D)$ from
\cite{witt} where the topology of $Isom (L(p,q))$ has been found in all 
cases (see Table ~\ref{tbl:i}). We see that $\fhom(\D)$ is indeed non-trivial
for all $L(p,q)$, including $S^3$, and hence these spaces are
candidates for anomalous theories.

\begin{table}[t]
\centering
\caption{ $\pi_1(\D(L(p,q)))$}
\vskip 1mm
\begin{tabular}{|c|c|c|}
\hline\hline

  $L(p,q)$       &  Topology of $Isom$ & $\fhom(\D)$  \\ \hline \hline

{$S^3$}  & $ \Z_2 \times S^3 \times \RP$ & $\Z_2$ \\
{$\RP$}  & $\Z_2 \times \RP \times \RP$ & $\Z_2 \oplus \Z_2$
\\
{$q^2\id 1\;{\rm mod}\;p$; $q\neq \pm 1 \;{\rm mod}\; p$}  & $\Z_2 \times \Z_2 \times S^1 \times
S^1$ & $\Z \oplus \Z$ \\ 
{$q^2\id -1\;{\rm mod}\;p$; $p \> 2$}  & $\Z_4 \times S^1 \times S^1$ & $\Z \oplus \Z$
\\
{$q\id \pm 1\;{\rm mod}\;p$; $p \> 2$}  & $\Z_2 \times S^1 \times \RP$ & $\Z \oplus \Z_2$
\\
{remaining cases}  & $\Z_2 \times S^1 \times S^1$ & $\Z \oplus \Z$
\\ \hline \hline
\end{tabular}
\label{tbl:i}
\end{table}
 
When the spatial manifold $\Sigma$ is asymptotically flat and of the form
$\re^3 \# L(p,q)$, the relevant group is $\Df(L(p,q))$, the frame fixing
diffeomorphism group of the one-point compactification, ${\bar
\Sigma}=L(p,q)$.  We can reduce the following exact sequence from
\cite{witt},
\begin{equation}
1 \rightarrow \fhom(\Df) \rightarrow \fhom(\D^+) \rightarrow \fhom(\Spa)\times
\Z_2  \rightarrow  {\pi}_0(\Df) {\buildrel a \over\rightarrow}
{\pi}_0(\D^+) \rightarrow 1,
\label{eqn:seq}
\end{equation}
(where $\D^+$ is the group of orientation preserving diffeomorphisms) to  
\begin{equation}
1 \rightarrow \fhom(\Df) \rightarrow \fhom(\D^+) \rightarrow \Z_p \oplus \Z_2
\rightarrow 1.
\label{eqn:red}
\end{equation}
To do this, we have used the isomorphism ${\pi}_0(\Df)\simeq
{\pi}_0(\D^+)$ for Lens spaces \cite{witt}, which implies that $a$ is an
isomorphism (since it is already surjective), and $\fhom(\Spa)= \Z_p$.
From (\ref{eqn:red}), $\fhom({\D}^+)/\fhom(\Df)=  \Z_p \oplus \Z_2$.
$\D^+$ and $\D$ differ only by orientation reversing diffeomorphisms
which are not identity connected, so that their first homotopy groups are
isomorphic. Using the right most column of Table ~\ref{tbl:i} we list $
\fhom(\Df)$ in Table  ~\ref{tbl:ii}.    

\begin{table}[t]
\centering
\caption{ $\pi_1(\D^+(L(p,q)))$}
\vskip 1mm
\begin{tabular}{|c|c|c|}
\hline\hline
$L(p,q)$    & $\fhom({\D}^+)/\fhom(\Df)=  \Z_p \oplus \Z_2$   &
$\fhom(\Df)$  \\ \hline \hline
{$S^3$}  & $\Z_2/\fhom(\Df)= \Z_2$ & $1$ \\      
{$\RP$}  & $(\Z_2\oplus \Z_2)/\fhom(\Df)= \Z_2 \oplus \Z_2$& $1$ \\      
{$q^2\id 1\;{\rm mod}\;p$; $q\neq \pm 1 \;{\rm mod}\; p$}  & $(\Z \oplus \Z)/\fhom(\Df)= \Z_p
\oplus \Z_2$ & $p\Z \oplus 2\Z$ \\      
{$q^2\id -1\;{\rm mod}\;p$; $p \> 2$}  & $(\Z \oplus \Z)/\fhom(\Df)= \Z_p
\oplus \Z_2$ & $p\Z \oplus 2\Z$ \\      
{$q\id \pm 1\;{\rm mod}\;p$; $p \> 2$}  &  $(\Z \oplus \Z_2)/\fhom(\Df)= \Z_p
\oplus \Z_2$ & $p\Z $\\      
{remaining cases}  & $(\Z \oplus \Z)/\fhom(\Df)= \Z_p
\oplus \Z_2$ & $p\Z \oplus 2\Z$ \\      
\hline \hline
\end{tabular}
\label{tbl:ii}
\end{table}

Again, we see that $\fhom(\D^+)$ is non-trivial for all $L(p,q)$ except
$\RP$ and $S^3$,  and hence are candidates for anomalous theories.

\subsection{Determining the Berry Phase}
Berry has shown that for a family of real Hamiltonians $\{\dham_q\}$, the
phase picked up by $\ho$ around a non-trivial loop $\gamma \subset
[\gamma]\in \fhom(\D(\Spa))$ is equal to $(-1)^k$, where $k$ is the number
of points in a disc $\disk$, bounded by $\gamma$, at which $\dham_q$ is
degenerate \cite{berry,simon}. In our case, $\gamma \subset \D$
parameterizes a family of real Dirac Hamiltonians and $\disk \subset
\R$. In the $\su$ case, the number of degeneracies $k$ was found by using
the Atiyah-Singer mod-2 index theorem for an appropriately constructed real
and antisymmetric 5-dimensional Dirac operator \cite{atisin,nelalv}. The
index theorem states that for such an operator, the number of zero modes is
a mod-2 invariant which was found by Witten to be odd in the $\su$ case
\cite{witten}. In the Hamiltonian framework \cite{nelalv}, the adiabatic
approximation was used to show that the kernel of the 5-dimensional Dirac
operator is related to the spectral flow of an associated 4-dimensional
operator, whose kernel is in turn related to the spectral flow of $\dham_q$
in $\disk$ and thence to the anomaly.
 
Unfortunately, in the case of gravity, a similarly constructed
5-dimensional Dirac operator is not real and hence the number of its zero
modes is {\it not} a mod-2 invariant. One might look for other ways to
calculate the number of degeneracies of $\dham_q$ in the disc, but here we
side step the issue a little by posing a slightly different question.

Namely, we consider gravity coupled to an ``external'' $\su$ gauge field
and ask whether this theory possesses an $\su$-induced diffeomorphism
anomaly. The existence of the external $\su$ gauge field ensures that the
associated 5-dimensional Dirac operator will be real. If a diffeomorphism
anomaly exists then it is ``induced'' by the $\su$ field. It is likely that
even for non-trivial spatial topologies, a pure $\su$ anomaly exits.  Our
concern in this paper however, will not be to determine whether the full
theory is anomalous or not, but to ascertain, rather, the role played by
the diffeomorphisms.

Let us consider the trivial $\su$ bundle $\su\times \Spa$ over $\Spa$.  The
configuration space for this theory is the Cartesian product, $\cA \times
\R$, where $\cA$ is the space of all 3-dimensional $\su$ connections, and
the full group of gauge transformations is $\G^3 \sd \D$. 
Note that since the $\su$ bundle is trivial, 
connection, $A^0\id 0 \in \cA $. Although the action of $\D\in \G^3 \sd \D$
leaves $A^0$ fixed, this is not true for a generic $A \in \cA$.

\begin{figure}[t]
\centerline{\epsfig{file=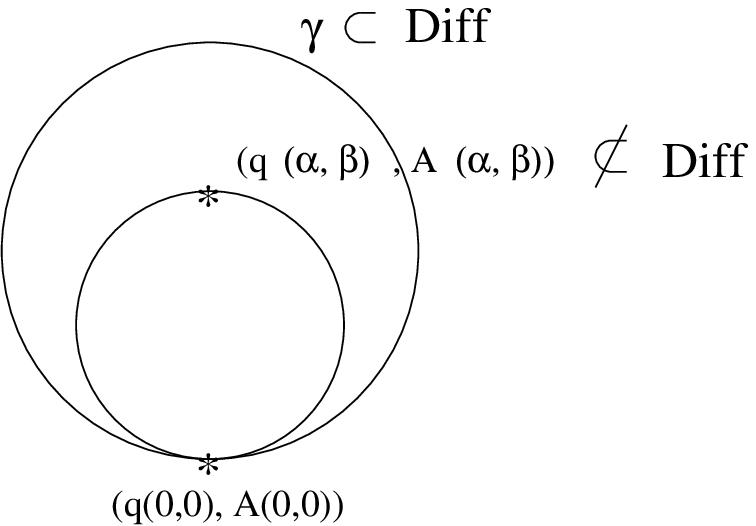,clip=3in,width=3in}}
\caption{Loops in \it{Riem}}
\label{fig:loop}
\end{figure}

Since we are only interested in the {\it diffeomorphism } anomaly, the
non-contractible loop we consider lies only in $\D \subset \G^3 \sd \D$.  Let
$\disk \subset \R$ be the disc suspended by $\gamma$ and let
$(A(\alpha,\beta),q(\alpha,\beta))$ be the $2$-parameter family of fields
that make up $\disk$ where $\alpha \in [0,1]$ and $\beta \in [0, 2\pi]$
(Figure ~\ref{fig:loop}). Let $( A(0,0),q(0,0))$ be the basepoint and $(
A(1,.),q(1,.))$ the fields along $\gamma$ ( where the notation $(\alpha,.)$
means that $\alpha$ is fixed and $\beta $ is allowed to vary.) Since
$\gamma \subset \D \subset \G^3\sd \D$, all points on it are purely gauge
related to each other and the spectrum of ${\dham}_q$ is invariant along
$\gamma$. Note that  ${\dham}_q$ is generically non-degenerate so that
starting from a non-degenerate basepoint, the spectrum along $\gamma$ will
be non-degenerate.

From the loop of 3-metrics $q(\alpha,.) \subset \R$, we construct the
Riemmanian 4-metric on $\Spa \times S^1$,
\begin{equation}
g(\alpha) =  \begin{pmatrix}  1 & 0\\
		0 &  q(\alpha,.)\\ 
\end{pmatrix}.
\label{eqn:blah}
\end{equation} 
which is periodic in $\beta$, and from the loop of 3-dimensional gauge
potentials $A(\alpha,.)$, we construct the 4-dimensional gauge potential,
\begin{equation} 
\A(\alpha)=\begin{pmatrix} 0\\
				A(\alpha,.)\\
\end{pmatrix},
\label{eqn:connection}
\end{equation} 
in the $\A_0\id 0$ gauge. For the purpose of brevity, we will henceforth omit
mentioning the $\su$ gauge field, whenever its role is obvious.

Varying $\alpha$, we get a $1$-parameter family of 4-dimensional
fields interpolating between $(\A(0),g(0))$ and
$(\A(1),g(1))$, where
\begin{equation} 
g(0)= \begin{pmatrix}  1 & 0\\ 0 & q(0,0)\\ 
\end{pmatrix} \qquad \& \qquad g(1)=\begin{pmatrix}  1 & 0\\ 0
& q(1,.)\\  
\end{pmatrix}.
\label{eqn:g1}
\end{equation} 
 and $\A(0)$ and $\A(1)$ are similarly fixed \footnote{ Note that  $g(0)$
is ``static'' since it has no $\beta$ dependence.}. 

Since the spatial metrics $q(1,.)$ are all diffeomorphism related to one
another and to $q(0,0)$ by a sequence of ``small'' diffeomorphisms
\footnote{ The loop $\gamma$ lies in the identity connected component of
$\D$.}, this sequence ``extends'' to to a single 4-dimensional
diffeomorphism on $\Spa \times S^1$. This is because any ``small'
diffeomorphism can be continuously deformed to the identity diffeomorphism;
in fact because $\D_0$ is a Lie group and thus has a manifold structure,
this deformation can be made smooth. These deformations thus take $\Spa
\times S^1$ smoothly into itself so that $g(0)$ and $g(1)$ are related by a
4-dimensional diffeomorphism. A 4-dimensional Dirac operator defined on
$g(0)$ would therefore have the same spectrum as one defined on $g(1)$. 

From the $1$-parameter family  $(\A(\alpha),g(\alpha))$, one can similarly
construct a single 5-dimensional Riemannian metric $G$ on $\Spa\times S^1
\times I$,
\begin{equation} 
G=    \begin{pmatrix}  1 & 0\\
			    0 &  g(.)\\ 
\end{pmatrix} 
\label{eqn:metric}
\end{equation} 
and a 5-dimensional gauge potential $\cAA$ (again in the $\cAA_0\id 0$ gauge.)

Now, consider a single loop of 3-metrics $q(\alpha,.)$.  As $\beta$ varies from
$0$ to $2\pi$, the spectrum of $\dham_q$ might undergo a spectral flow,
although it must return to itself. Since the energy spectrum of $\dham_q$
consists of pairs $(E, -E)$, where $E$ is real, a spectral flow leads to a
cross-over point ${\beta}_i$ at which $\dham_{q(i)}$ is degenerate. Thus,
the loop $q(\alpha, .)$ , will contain as many degenerate points  ${\beta}_i$
as its spectral flows.

In order to find the number of these spectral flows, consider the Euclidean
Dirac operator $\;i \dfo$ on $\Spa \times \re$ with the  Riemannian
4-metric $g(\alpha)$ (\ref{eqn:blah}) and connection $\A(\alpha)$  (\ref{eqn:connection}),  whose induced spatial metric on $\Spa$ varies
adiabatically with ``time'' $t$ along $\gamma$ (as $t
\rightarrow \pm \infty$, the induced metric on $\Spa$ goes to ${q}(0,0)$.) 
Assuming the adiabatic approximation \cite{witten,nelalv}, the zero mode
equation $\;i \dfo \psi(t, x)=0$ reduces to
\begin{equation} 
{\partial}_{t}\psi(t, x)\approx -{\gamma}^4{\gamma}^iD_i\psi(t,x),
\end{equation}
where $x$ labels the spatial coordinates.  Since
$[{\partial}_{t},{\gamma}^4{\gamma}^iD_i]\approx 0$, these modes can be
separated into $\psi(t, x)=G(t)\chi(x)$ where $\chi(x)$ satisfies the
eigenvalue equation, ${\gamma}^4{\gamma}^iD_i\chi(x) = E(t)\chi(x)$. Thus,
\begin{equation} 
G(t)=G(0)e^{-\int{E(t')dt'}}.
\end{equation}
Now, the spinor $\psi$ on $\Spa \times \re$ will not in general be
normalisable.  However,  when $G(t)$ {\it is} normalisable, 
as $t\rightarrow - \infty$, $E(t)$ is negative and  as
$t\rightarrow + \infty$, $E(t)$ is positive. Thus, although the
spectrum of ${\gamma}^4{\gamma}^iD_i$ is the same as $t \rightarrow \pm
\infty$, there will be a spectral flow for every normalisable zero mode of
$\;i \dfo$. The spectrum  of ${\gamma}^4{\gamma}^iD_i$ coincides with that of
${\dham}_q(t)={\gamma}^0{\gamma}^iD_i$, so that a spectral flow for
${\gamma}^4{\gamma}^iD_i$ is a spectral flow for ${\dham}_q(t)$.

In restricting to normalisable zero modes of $\;i \dfo$, one is effectively
compactifying $\Spa \times \re$ to $\Spa \times S^1$, since $\psi
\rightarrow 0$ as $t \rightarrow \pm \infty$, and $\Spa$ is ``essentially''
closed (i.e., it is not a one-point compactification). Since the fields on
$\Spa \times  \re$ are periodic, the trivial $\su$ bundle on $\Spa \times
\re$ is compactified to a trivial bundle over  $\Spa \times
S^1$. The associated Dirac operators are then identical except for
the existence of non-normalisable modes in the latter case.
Thus, the number of spectral flows of $\dham$ along the loop
$q(\alpha,.)$ = the number of zero modes of the associated $\;i \dfo$ on $\Spa
\times S^1$. However, there is no obvious method of finding the number of
zero modes of $\;i \dfo$ at this stage.

Let $\; i \diD$ be a Dirac operator on $\Spa \times S^1 \times \re$ with
the Riemannian 5-metric $G$ (\ref{eqn:metric}) whose induced metric on
$\Spa \times S^1$ varies adiabatically with another ``time'' parameter
$\tau$, from $g(0)$ to $g(1)$, defined in (\ref{eqn:g1}). Using the same
adiabatic approximation technique as above for the zero mode equation $\; i
\diD \Psi(\tau,t,x) =0$, we get
\begin{equation} 
{\partial}_{\tau}\Psi(\tau,t,x)=
-{\gamma}^5{\gamma}^{\mu}{{\mathfrak D}^4}_{\mu}\Psi(\tau,t,x). 
\end{equation}
Again, the number of normalisable zero modes gives of $\; i \diD$ = the
number of spectral flows of ${\gamma}^5\;\dfo$, which has the same spectrum
as $i\;\dfo$.  The eigenvalues of $i\;\dfo$ come in pairs, $(\lambda,
-\lambda)$ so that a spectral flow leads to a cross-over point ${\alpha}_i$
at which $i\;\dfo({\alpha}_i)$ has a pair of non-trivial zero modes.  Since
we are only interested in chiral fermions, only one of these zero modes is
relevant to us. Now, a single zero mode of $\;i \dfo({\alpha}_i)$ gives a
single spectral flow of $\dham({\alpha}, {\beta})$ along $q(\alpha, .)$ and
hence a single degeneracy along this loop at $({\alpha}_i,
{\beta}_j)$. Thus, the number of normalisable zero modes of $\; i \diD$ =
number of degeneracies in $\disk$. The logic of this construction should be
clear by now.

Again, it would seem that the restriction to normalisable zero modes of $\;
i \diD$ would allow the compactification $\Spa \times S^1 \times \re$ to
$\Spa \times S^1 \times S^1$.  However, the fields on $\Spa \times S^1
\times \re$ are not periodic. In the pure $\su$ case in \cite{witten}, for
example, a non-trivial $\Z_2$ gauge transformation between the connections
at $t \rightarrow \pm \infty$ gives a non-trivial $\su$ bundle over $S^5$
under compactification. In our case, however, since we are dealing with
only the diffeomorphisms, such a $\Z_2$ twist in the $\su$ gauge potentials
cannot appear, and the relevant compactification leads to a trivial $\su$
bundle over $\Spa \times S^1 \times S^1$. The non-periodicity in the metric
on the other hand is only the statement that $G$ cannot be deformed
continuously to a metric constant in $\tau$.

Now,
\begin{equation} 
\diD= ({\gamma}^M \otimes 1)(1{\nabla}^5_M \otimes 1 + i1
\otimes \cAA_M^a T^a),
\label{eqn:dirac}
\end{equation}
where $\nabla^5$ is the 5-dimensional gravitational connection compatible
with $G$, ${\gamma}^M$ is the psuedoreal representation of the generators
of the Clifford algebra ${{\cal C}}(0,5)$ \cite{brumor}, and $T^a$ are a
$2$-dimensional psuedoreal representation of  the generators of the {\bf
su(2)} Lie algebra. Since the Clifford algebra has 4 spinor
dimensions and the representation of the {\bf su(2)} Lie
algebra is  $2$-dimensional, $\diD$ acts on an $8$-dimensional spinor space. 

Since $\diD$ is a real antisymmetric operator on a compact manifold, the
Atiyah-Singer index theorem \cite{atisin} tells us that the number of its
zero modes are a mod-2 invariant. In other words, given the number $n$ of
zero modes for {\it any } $\di$ on $\Spa \times S^1\times S^1$, the number
of zero modes of $\diD$ mod 2 = $n$ mod 2, provided $(G', \cAA')$ can be
continuously deformed to $(G, \cAA)$.  This is precisely what we need to
determine the anomaly, since it only depends on the number of degeneracies
mod 2 in $\disk$. Thus our problem reduces to finding $n$ for a suitable
pair of fields $(G', \cAA')$ on the trivial $\su$ bundle over $\Spa \times S^1\times S^1$.


Since the $\su$ bundle is trivial, an immediate choice for the connection
is $\cAA_{M}'\id 0$. The zero mode equation reduces to,
\begin{equation}
\di \Psi =
({\snabla}^5 \otimes \1)({\Psi}_{(1)}\otimes {\Psi}_{(2)}) =0,
\end{equation}
where $\Psi =({\Psi}_{(1)}\otimes {\Psi}_{(2)})$ is a tensor product of a 4
and a 2-dimensional spinor.  Thus, for any zero mode $\Psi$ of $\di$,
${\Psi}_{(1)}$ must be a zero mode of ${\snabla}^5$, i.e.,
${\snabla}^5{\Psi}_{(1)}=0$. Now, if $\{{\Psi}_{(1)}^i\}$ is the set of
zero modes of $\gdi$ then $\Psi={\Psi}_{(1)}^i\otimes {\Psi}_{(2)}$ is a
zero mode of $\di$ for {\it any} normalisable 2-spinor ${\Psi}_{(2)}$. On
the other hand, if $\gdi$ has {\it no} zero modes, then $\di$ too has no
zero modes. Below, we employ suitable metrics on $S^3$ and the Lens spaces
$L(p,q)$ for which $\gdi$ has no zero modes. This small but crucial point
helps us establish unambiguously that the number of zero modes of $\diD$ is
even, which $\Rightarrow$ there are an even number of degeneracies in
$\disk$ and therefore no anomaly.

\subsection{Zero Modes of the Dirac operator }
We now calculate the zero modes of $\gdi$ on $S^3 \times S^1 \times S^1$
with the metric $ {\rd}s^2 = \rd \tau^2 + \rd t^2 + {\rd}^2{\Omega},$ where
$\tau$ and $t$ parameterize the two $S^1$'s and $\rd^2\Omega= \rd^2 \psi +
\sin^2\psi(\rd^2\theta +\sin^2\theta \rd^2\psi)$ is the standard metric on
$S^3$.

We start with the 4-dimensional space $S^3 \times S^1$ with induced
metric $\rd t^2 + {\rd}^2{\Omega}$.  The Dirac operator $\gdfo$ on this
space  satisfies,
\begin{equation} 
\gdfo \;\psi = ({\gamma}^4 {\partial}_{t} + {\gamma}^{i} {\tnabla}_i)
\;\psi = i \lambda \;\psi.
\label{eigen}
\end{equation}
where the $\{\gamma^{\mu}\}$ are the chiral representation of the
associated Clifford algebra,
\begin{equation} 
{\gamma}^4 = \begin{pmatrix}  0 & 1\\  1 &
0\\ \end{pmatrix} \quad \& \quad  {\gamma}^m= \begin{pmatrix} 0 &
i{\sigma}^m \\ -i{\sigma}^m & 0 \\ \end{pmatrix},
\end{equation}
and ${\tnabla}$ is the 3-dimensional connection on $S^3$.  Decomposing
$\psi$ into the two $2$-dimensional spinors, ${\phi}_{\pm}$, $\psi = \left
[\begin{matrix} {\phi}_+ \\ {\phi}_- \\
\end{matrix}\right]$,  we obtain the two coupled
first-order differential equations,
\begin{eqnarray}
\1{\partial}_{t} {\phi}_- + i {\sigma}^m {\widetilde e}_m^i {\tilde
\tnabla}_i {\phi}_- & = & i \1 \lambda {\phi}_+ \\ 
\1{\partial}_{t} {\phi}_+ - i
{\sigma}^m {\widetilde e}_m^i {\tilde \tnabla}_i {\phi}_+ & =&  i \1 \lambda
{\phi}_-,
\end{eqnarray}
where  ${\widetilde e}_m^i$ and ${\tilde \tnabla}_i$ are respectively,
the triad and standard connection  on $S^3$. Or,
\begin{equation}
(\1 {{\partial}_{t}}^2 + ({\tilde \tsnabla})^2){\phi}_{\pm}= -\1
{\lambda}^2{\phi}_{\pm},
\label{laplacian}
\end{equation}
where $i{\tilde \tsnabla}={\sigma}^i{\tilde \tnabla}_i$ is the
3-dimensional Dirac operator on $S^3$.  Since $[
{{\partial}_{t}}^2,({\tilde \tsnabla})^2]=0$, we can decompose
${\phi}_{\pm} = N(t)^{\pm}{\chi}_{(n\pm)} (\Omega)$, with
${\chi}_{(n\pm)}(\Omega)$ satisfying $\tilde \tsnabla
{\chi}_{(n\pm)}(\Omega)= i {\mu}^{\pm} {\chi}_n(\Omega) $ and ${\mu}^{\pm}$
being real. From reference \cite{camhig}, we find that ${\mu}^{\pm}=\pm (n
+ {3 \over 2})$, where $n$ is a positive integer. Hence $\tilde \tsnabla$
has no zero modes on $S^3$ with the standard metric $\rd^2 \Omega$.
Equation (\ref{laplacian}) then reduces to
\begin{equation}
{\partial}_{t}^2N(t)= - ({\lambda}^2 - {\mu}^2)N(t)= -l^2 N(t).
\end{equation}
and since ${\partial}_{t}^2$ is an operator on $S^1$, its eigenfunctions
must therefore satisfy the periodicity condition $l \in \Z$. $\lambda$
in turn  satisfies
\begin{equation} 
{\lambda}^2 = l^2 + {\mu}^2 = l^2 + {(n + {3\over 2})}^2,
\label{eqn:lam}
\end{equation}
so that
$|\lambda|_{min} ={9 \over 4}$. Thus $\gdfo$ has no
zero modes on  $S^3
\times S^1$ with  metric $ \rd s^2 = \rd t^2 + {\rd} ^2{\Omega}$.   

For the 5-dimensional operator ${\snabla}^5$ on $S^3 \times S^1 \times S^1$, the
Clifford algebra includes 
\begin{equation} 
{\gamma}^5=\begin{pmatrix}  1 & 0\\  0 & -1\\ \end{pmatrix}. 
\end{equation}
Let
\begin{equation} 
\gdi \;\Psi = ({\gamma}^5{\partial}_{{\tau}}+\gdfo)\;\Psi= ik\;\Psi,
\end{equation} 
or,
\begin{equation} 
({{\partial}_{{\tau}}}^2+(\gdfo)^2)\;\Psi= - k^2\;\Psi.
\end{equation}
Since $[{{\partial}_{{\tau}}}^2, (\gdfo)^2]=0$, $\Psi$ can be expressed as 
$\Psi=R({\tau})\psi(t, \Omega)$, where $(\gdfo)^2\;\psi(t,\Omega)=
-{\lambda}_{n, l}^2 \;\psi$ and ${\lambda}_{n, l}^2= l^2 + {(n + {3\over
2})}^2$ (\ref{eqn:lam}). Now, ${{\partial}_{{\tau}}}^2 R({\tau})=
-r^2 R({\tau})$, where $r \in \Z$, so that, 
\begin{equation} k^2 = r^2 +{\lambda}^2=  r^2 + l^2 +({n + {3\over
2}})^2. 
\label{eqn:k}
\end{equation}  
Hence ${k^2}_{min}= {9\over 4}$ and we can conclude that $\gdi$ too has no
zero modes for the metric $ {\rd}s^2 = {\rd}{\tau}^2 +\rd t^2 +
{\rd}^2{\Omega},$ on $S^3 \times S^1 \times S^1$.

Although this calculation has been done for a specific 5-dimensional metric
the mod 2 index theorem for $\di$ implies that the number of zero modes
will only change by $2$ if one moves continuously to another point in the
space of fields on $S^3 \times S^1 \times S^1$.  Thus the number of zero
modes of $\diD$ (\ref{eqn:dirac}) must be even. Putting this information
into our previous arguments, we see that there are no diffeomorphism
anomalies when the spatial manifold is $S^3$.

This can be generalised to the case of the other Lens spaces by changing
the 3-dimensional metric ${\rd}^2{\Omega}$ on $S^3$ to an appropriate one
${\rd}^2{\Omega}_L$ on $L(p,q)$, so that $ {\rd}s^2 = {\rd}{\tau}^2 +\rd
t^2 + {\rd}^2{\Omega}_L$ is the metric on $L(p,q) \times S^1 \times
S^1$. Repeating the above procedure, we again find the relation
(\ref{eqn:lam}) where $i \mu$ is now replaced by the eigenvalue of $\tilde
\tsnabla$ on $L(p.q)$ with metric ${\rd}^2{\Omega}_L$. If $|\mu|_{min}\neq
0$ then from (\ref{eqn:k}) it is immediately obvious there are no zero
modes for the associated $\gdi$ in this case as well.

Now, heuristically, it can be argued that since $L(p,q)$ is the quotient
manifold $S^3 / \Z_p$, its Dirac operator must have a spectrum smaller than
the spectrum on $S^3$ whenever $\rd^2 \Omega_L$ is a ``$\Z_p$'' quotient of the
standard metric $\rd^2\Omega$ on $S^3$, and so doesn't possess zero modes
either.

The spectrum of the Dirac operator $\tsnabla$ has in fact been explicitly
calculated in reference \cite{bar}, using the so-called ``Berger''
metrics. Fixing the parameter $T$ that appears there to $1$ gives us
metrics $\rd^2 \Omega_L$ on $L(p,q)$ which are effectively ``quotient''
metrics of $\rd^2\Omega$. The results from \cite{bar} show that the
spectrum of $\tsnabla$ indeed does not possess zero modes.  Thus there are
no diffeomorphism anomalies for $\Sigma =L(p,q)$ either.

\section{Conclusions}
We have established that for $3+1$ quantum gravity, assuming the validity
of the strong form of the Hatcher conjecture, $\su$-induced global
diffeomorphism anomalies are absent whenever the spatial slice is
diffeomorphic to $S^3$ or any Lens space $L(p,q)$. In the case of pure
gravity, this question is still remains open. We briefly explore the
implications of our result.

For a generic 3-manifold, there exists an infinity of inequivalent quantum
sectors of 3+1 canonical gravity \cite{sumati}. This feature is clearly not
desirable in a fundamental theory and it is believed that higher energy
effects like topology change would be necessary to resolve this problem.
If the ``frozen'' topology theory is viewed as an effective description,
however, one might hope that anomalies would provide a selection rule for
the allowed spatial topology. Namely, those manifolds which lead to an
infinite number of sectors would be ``anomalous'' (in the sense that the
quantum theory on these manifolds possesses non-perturbative global
diffeomorphism anomaly of the kind we have investigated .)  We notice that
the number of quantum sectors is in fact finite when $\Sigma = L(p,q)$ and
thus, in a sense, one doesn't need an anomaly to ``rule'' out these spatial
topologies. This is in keeping with our result, as stated above.


It is generally believed that the full theory of quantum gravity will also
be capable of describing processes involving topology change.  In the low
energy effective description of this full theory, topology change to
manifolds that are ``anomalous''  is expected to be highly suppressed. Our
result indicates that all Lens spaces are ``well-behaved'', and that
topology change involving these spaces will not be forbidden.

\noindent{\bf Note added:} After completion of our work we became aware of
a paper by Chang and Soo \cite{chasoo} where the question of global
anomalies in the self-dual formulation of gravity is discussed using
generalised spin structures. The results obtained by these authors have no
implications on ours, as they discuss anomalies in a specific formulation
of canonical gravity (namely the self-dual formulation) and for the
topology of spacetime being $S^4$.

\section {Acknowledgments}
We would like to thank A. P. Balachandran for guiding us through problem,
and Don Marolf, Petre Golombeanau and Alan Steif for discussions. We are
also grateful to E. Witten for explaining the construction of the
five-dimensional instanton in \cite{witten} and L. Alvarez-Gaum\'e for
explaining the Berry phase argument used in \cite{nelalv}. The research of
S. S. was supported by a Syracuse University Fellowship, and that of
S. V. supported in part by U. S. DOE grant under contract
no. DE-FG02-85ER40231. 

\bibliographystyle{unsrt}

\bibliography{gr_anomaly}

\begin{thebibliography}{10}

\bibitem{witten}
E.~Witten.
\newblock An {SU(2)} anomaly.
\newblock {\em Phys. Lett.}, B117:324--328, 1982.

\bibitem{nelalv}
P.~Nelson and L.~Alvarez-Gaum\'e.
\newblock Hamiltonian interpretation of anomalies.
\newblock {\em Commun. Math. Phys.}, 99:103--114, 1985.

\bibitem{segal}
Graeme Segal.
\newblock Fadeev's anomaly in {G}auss' law.
\newblock Oxford University Preprint.

\bibitem{witt}
D.~M. Witt.
\newblock Symmetry groups of state vectors in canonical quantum gravity.
\newblock {\em J. Math. Phys.}, 27:573--592, 1986.

\bibitem{hatcher1}
A.~Hatcher.
\newblock Linearization in $3$-dimensional topology.
\newblock {\em Proc. Int. Congr. Math. Helsinki}, pages 463--468, 1978.

\bibitem{chaerc}
L.~Chandar and E.~Ercolessi.
\newblock Inequivalent quantizations of {Y}ang-{M}ills theory on a cylinder.
\newblock {\em Nucl. Phys.}, B(426):94--106, 1994.

\bibitem{bamalfi}
A.P. Balachandran.
\newblock {Classical Topology and Quantum Phases: {Q}uantum {M}echanics}.
\newblock In S.~de~Filippo, M.~Marinaro, and G.~Marmo, editors, {\em
  Geometrical and Algebraic Aspects of Nonlinear Field Theories}, pages 1--28,
  Amalfi, Italy, May 1988. Elsevier, Amsterdam, 1989.

\bibitem{atisin}
M.~F. Atiyah and I.~M. Singer.
\newblock The index of elliptic operators:{V}.
\newblock {\em Ann. of Math.}, 93:139, 1971.

\bibitem{hatcher2}
A.~Hatcher.
\newblock A proof of a {S}male conjecture, ${D}iff({S}^{3}) \simeq {O}(4)$.
\newblock {\em Ann. of Math.}, 117(3):553--607, 1983.

\bibitem{ivanov}
N.~V. Ivanov.
\newblock Homotopies of automorphism spaces of some three-dimensional
  manifolds.
\newblock {\em Sov. Math. Dokl.}, 20(1):47--50, 1979.

\bibitem{rubinstein}
J.~H. Rubinstein.
\newblock On $3$-manifolds that have finite fundamental group and contain
  {K}lein bottles.
\newblock {\em Trans. Am. Math. Soc.}, 251:129--137, 1979.

\bibitem{rubbir}
J.~H. Rubinstein and J.~S. Birman.
\newblock One-sided {H}eegaard splittings and homotopy groups of some
  $3$-manifolds.
\newblock {\em Proc. London Math. Soc.}, 49(3):517--536, 1984.

\bibitem{hodrub}
C.~J. Hodgeson and J.~H. Rubinstein.
\newblock Involutions and isotopies of {L}ens spaces.
\newblock In {\em Knot theory and manifolds (Vancouver, B.C., 1983)}, Lecture
  Notes in Math.,1144, pages 60--96, New York, 1985. Springer.

\bibitem{berry}
M.~V. Berry.
\newblock Quantal phase factors accompanying adiabatic changes.
\newblock {\em Proc. R. Soc. London Ser. A}, 392:45--57, 1984.

\bibitem{simon}
Barry Simon.
\newblock Holonomy, the quantum adiabatic theorem, and {B}erry's phase.
\newblock {\em Phys.Rev.Lett}, 51:2167--2170, 1983.

\bibitem{brumor}
Y.~Choquet-Bruhat and C.~DeWitt-Morette.
\newblock {\em Analysis, Manifolds and Physics, Part II: 92 Applications}.
\newblock North-Holland, New York, 1989.

\bibitem{camhig}
R.~Camporesi and A.~Higuchi.
\newblock On the eigenfunctions of the {D}irac operator on spheres and real
  hyperbolic spaces.
\newblock e-Print Archive: gr-qc/9505009.

\bibitem{bar}
Christian Bar.
\newblock The {D}irac operator on homogeneous spaces and its spectrum on
  3-dimensional lens spaces.
\newblock {\em Arch.Math.}, 59:65--79, 1992.

\bibitem{sumati}
Rafael Sorkin and Sumati Surya.
\newblock An analysis of the representations of the mapping class group of a
  multi-geon three manifold.
\newblock e-Print Archive: gr-qc/9605050.

\bibitem{chasoo}
Lay~Nam Chang and Chopin Soo.
\newblock The standard model with gravity couplings.
\newblock {\em Phys. Rev. D}, 53:5682--5691, 1996.

\end{thebibliography}
\end{document}